# Broadband Ultrafast Dynamics of Refractory Metals: TiN and ZrN

*Benjamin T. Diroll,[1]\* Soham Saha,[2] Vladimir M. Shalaev,[2] Alexandra Boltasseva,[2] and Richard D. Schaller[1,3]*

[1]Center for Nanoscale Materials, Argonne National Laboratory, Lemont, Illinois 60439

[2]School of Electrical and Computer Engineering, Purdue University, West Lafayette, Indiana 47907

[3]Department of Chemistry, Northwestern University, 2145 Sheridan Road, Evanston, Illinois 60208

\*bdiroll@anl.gov

Abstract:

Transition metal nitrides have recently gained attention in the fields of plasmonics, plasmon-enhanced photocatalysis, photothermal applications, and nonlinear optics because of their suitable optical properties, refractory nature, and large laser damage thresholds. This work reports comparative studies of the transient response of films of titanium nitride, zirconium nitride, and Au under femtosecond excitation. Broadband transient optical characterization helps to adjudicate earlier, somewhat inconsistent reports regarding hot electron lifetimes based upon single wavelength measurements. These pump-probe experiments show sub-picosecond transient dynamics only within the epsilon-near-zero window of the refractory metals. The dynamics are dominated by photoinduced interband transitions resulting from ultrafast electron energy redistribution. The enhanced reflection modulation in the epsilon-near-zero window makes it possible to observe the ultrafast optical response of these films at low pump fluences. These results indicate that electron-phonon coupling in TiN and ZrN is 25-100 times greater than in Au. Strong electron-phonon coupling drives the sub-picosecond optical response and facilitates greater lattice heating compared to Au, making TiN and ZrN promising for photothermal applications. The spectral response and dynamics of TiN and ZrN are only weakly sensitive to pump fluence and pump excitation energy. However, the magnitude of the response is much greater at higher pump photon energies and higher fluences, reaching peak observed values of 15 % in TiN and 50 % in ZrN in the epsilon-near-zero window.

1. Introduction

Metallic nitride materials such as titanium and zirconium nitrides have a long history as conducting and durable coatings. Recent work on transition metal nitrides has highlighted their optical properties as alternatives to noble metals employed in plasmonics such as gold (Au).[1–9] Titanium nitride (TiN), zirconium nitride (ZrN), and other plasmonic nitrides such as hafnium nitride (HfN) and tantalum nitride (TaN) are particularly attractive due to their high melting points that bolster stability at higher ambient temperatures[10–12] and/or under higher laser irradiation intensities,[13–16] in addition to their mechanical hardness[17,18] and complementary metal-oxide-semiconductor (CMOS) compatibility.[19–21] Recent work has demonstrated that TiN shows strong local heating compared to Au[22–24] which may be exploited for photothermal therapy,[25,26] shape-memory effects,[27] thermochromic windows,[28] photoreactions,[29–32] heat transducers or thermophotovoltaic materials,[22,33–37] or photodetection.[38] Implicit in these observations and devices are very different optical responses of metallic nitrides compared to gold—the most similar classical plasmonic material—particularly with regard to the dissipation of heat.

Transient reflection spectroscopy offers a window into the dynamics of heating in metal nitrides upon photoexcitation. Previous pump-probe spectroscopy on titanium nitride and other refractory metals has mainly consisted of single pump and probe wavelengths.[39–42] While some broadband transient absorption studies have been performed on these materials,[43] the epsilon-near-zero (ENZ) window where the real part of the dielectric permittivity crosses zero, has thus far remained unexplored. Importantly, based upon earlier spectroscopic investigations of noble metals and heavily-doped semiconductors, the largest changes of transient reflectivity are anticipated at ENZ regions.[44–47] Furthermore, earlier works yield contradictory conclusions from ultrafast pump-probe studies. Early data, in which sub-picosecond dynamics were not observed, suggested that electron-phonon coupling in TiN is weak,[39] with electron equilibration with the lattice requiring tens or hundreds of picoseconds in comparison to ~1 ps in gold.[48–51] Subsequent experiments with high pump fluences were able to observe the fast dynamics, conveying strong electron-phonon coupling resulting in sub-picosecond equilibration of electrons and the lattice.[42]

In this work, high quality films of TiN and ZrN grown on magnesium oxide (MgO) substrates are examined by broadband transient reflectivity measurements at visible and ultraviolet energies. These results are compared with the static differential reflectivity of the same films and transient reflectivity of Au films on silicon. Under a range of pump photon energies and fluences, transient reflectivity measurements of the metallic nitrides show a sub-picosecond transient dynamic in the ENZ region attributed to photoinduced interband transitions enabled by transient heating and cooling of the electron plasma. Fitting of the reflectivity dynamics permits an estimation of the electron-phonon coupling constants which are 25-100 times larger than Au. Strong electron-phonon coupling means that lattice heating yields substantially greater contributions to the optical response compared to other plasmonic metals like Au or Ag. This strong electron-phonon coupling marks TiN and ZrN as particularly promising for photothermal applications. Comprehensive measurements show that the spectral response and dynamics are only weakly sensitive to the pump fluence and pump photon energy, but the amplitude of reflection modulation increases substantially with higher pump photon energies and higher fluence, reaching peak observed modulations of 15 % and 50 % in TiN and ZrN with 7.33 mJ·cm$^{-2}$ of 3.10 eV excitation, respectively.

2. Static Properties of Nitride Films

The static optical properties of the studied samples are shown in Figure 1. The optically thick films (200 nm) of TiN and ZrN are grown on magnesium oxide (MgO). Compared to other substrates, those grown on MgO show suppressed optical losses.[3] Scanning electron microscopy (Figure S1) and atomic force microscopy (Figure S2) show that the films have a roughness much smaller than the optical wavelengths used in these experiments, with ZrN showing somewhat higher roughness than TiN. Figure 1a shows the normal-incidence reflectivity of the samples, as well as that of a gold film (120 nm with 2 nm chromium adhesion layer) on silicon. The reflectivity of the TiN and ZrN samples in Figure 1a approaches unity at low energy and dips to a minimum at higher energy, where interband transitions occur.[52–54] The results of spectroscopic ellipsometry, which are similar to earlier reports,[4,55] are shown in Figure 1b. Both TiN and ZrN are modeled using a common approach of a Drude oscillator at low energy and Lorentz oscillators at higher energy to represent interband transitions (See Experimental Methods). Based upon calculated band structure[56,57] the interband transitions are assigned to transitions from *p* states below the Fermi energy to *d* bands above the Fermi energy,[54] although not all reports convey an identical interpretation.[52,53] These nitride films show a zero crossing of the real permittivity at 2.63 eV and 3.18 eV for TiN and ZrN, respectively. This ENZ region is particularly noted as changes in the electron plasma or lattice temperature of the materials are anticipated to generate the largest relative changes in reflectance where $\varepsilon'$ is small.[44,46,58–60]

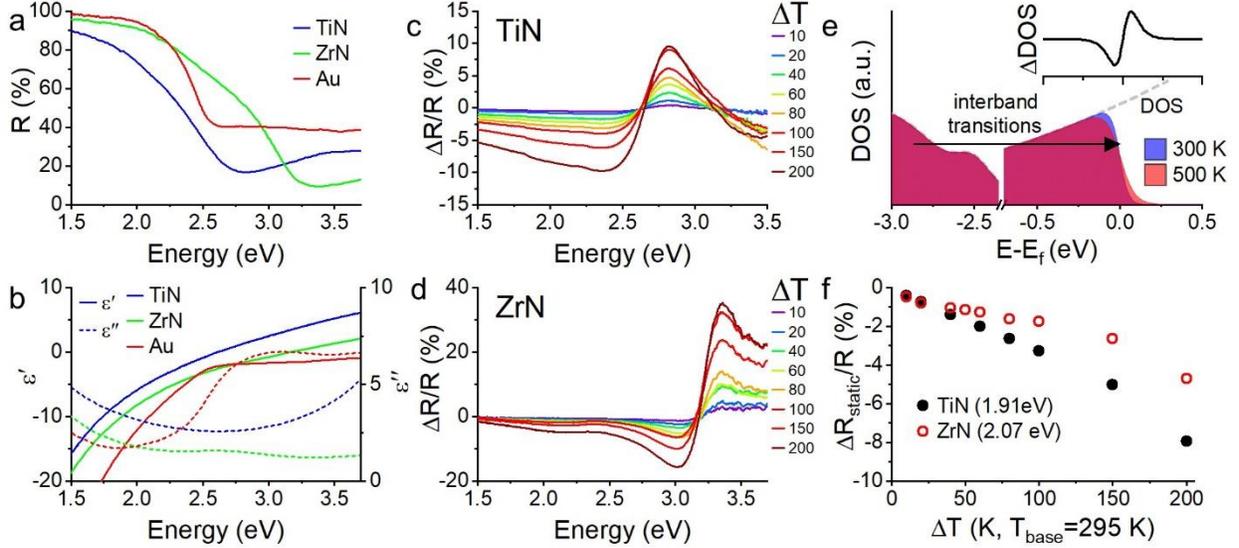

Figure 1. (a) Normal-incidence specular reflectivity of TiN and ZrN films grown on MgO substrate and a Au film on silicon. (b) Dielectric functions of TiN, ZrN (both on MgO) and Au films. The real part of the dielectric function is shown in solid lines and the imaginary part is shown in dashed lines. Static differential reflectance data collected for (c) TiN and (d) ZrN films. Differential spectra are calculated from initial reflectance at 295 K. (e) Model of the Fermi-Dirac probability distribution convolved with a cartoon density of states (DOS) distribution loosely based upon calculated TiN band structure.[56] Interband transitions occur from p to d bands.[54,56,57] Inset shows the change in occupation of the density of states with temperature change. (f) Scatterplot of the differential reflectivity versus temperature difference at 295 K.

To understand the contribution of heating on TiN and ZrN films upon photoexcitation, measurements of the sample reflectivity with elevated temperature are presented first. The fractional change in reflectivity ($\Delta R/R$) for the samples as a function of the change in temperature ($\Delta T$) above room temperature (295 K) are shown in Figures 1c and 1d. With temperature elevation, the samples show an increase in reflectivity at higher energies than the ENZ point and a decrease of reflectivity at lower energies. (A similar line-shape is observed Au.) Particularly noteworthy is that the ZrN sample shows much larger differences in reflectivity near the ENZ than the TiN sample for the same change in temperature. The static $\Delta R/R$ spectra chiefly reflect changes of allowed interband transitions, but also includes contributions from electron-phonon coupling due to thermal motion and the changing energy of all transitions upon lattice expansion.[10,61–64] The dominant role of interband transitions in the static reflectivity changes of TiN and ZrN are shared with Au. Elevation of the equilibrium temperature broadens the Fermi-Dirac distribution, leading to increased occupation of states above the Fermi level and depopulation of states below the Fermi level. Figure 1e illustrates this effect with a cartoon density of states modeled on TiN and ZrN.[4,56,57] The change in the electron energy distribution (shown inset on Figure 1e) translates into changes the spectral distribution of available interband transitions, resulting in the static thermal difference reflectivity spectra as in Figures 1c and 1d.[61–63]

Static differential reflectivity data establishes baseline expectations for the change in reflectivity of the samples (under equilibrium conditions) and offers an empirical calibration for estimation of the film lattice temperature following photoexcitation.[65–68] In typical thermoreflectance measurements, reflectivity is considered to vary linearly with temperature according to a thermoreflectance coefficient $C_{TR}(E)$:

$$\frac{\Delta R(E)}{R(E)} = C_{TR}(E)\Delta T \quad (1)$$

For calibration purposes, the reflectivity of the samples in the metallic region is used because it is not strongly affected by different spectral resolutions of the transient and static measurements or ultrafast dynamics in the ENZ region. Figure 1f shows that ΔR/R is linear in temperature up to ΔT of 150 K in the metallic region, with the $C_{TR}$ coefficient for TiN estimated at -3.321×10$^{-4}$·K$^{-1}$ (at 1.91 eV or 650 nm) and for ZrN, -2.055×10$^{-4}$·K$^{-1}$ (at 2.07 eV or 600 nm). With greater temperature elevation, real permittivity in the metallic region increases nonlinearly,[10] which is most likely responsible for the deviation of linear reflectivity change at higher temperatures.

3. Overview of Transient Reflectivity

Broadband transient reflectivity maps of the TiN, ZrN, and Au films are shown in Figure 2. The maps convey common elements between the refractory metals and Au, but they also show temporal dynamics that are dramatically different at both short (< 1 ps) and long (> 10 ps) delay times. For all samples, photoexcitation increases reflectivity on the blue side of the ENZ and decreases reflectivity in the metallic region, like static reflectivity changes.

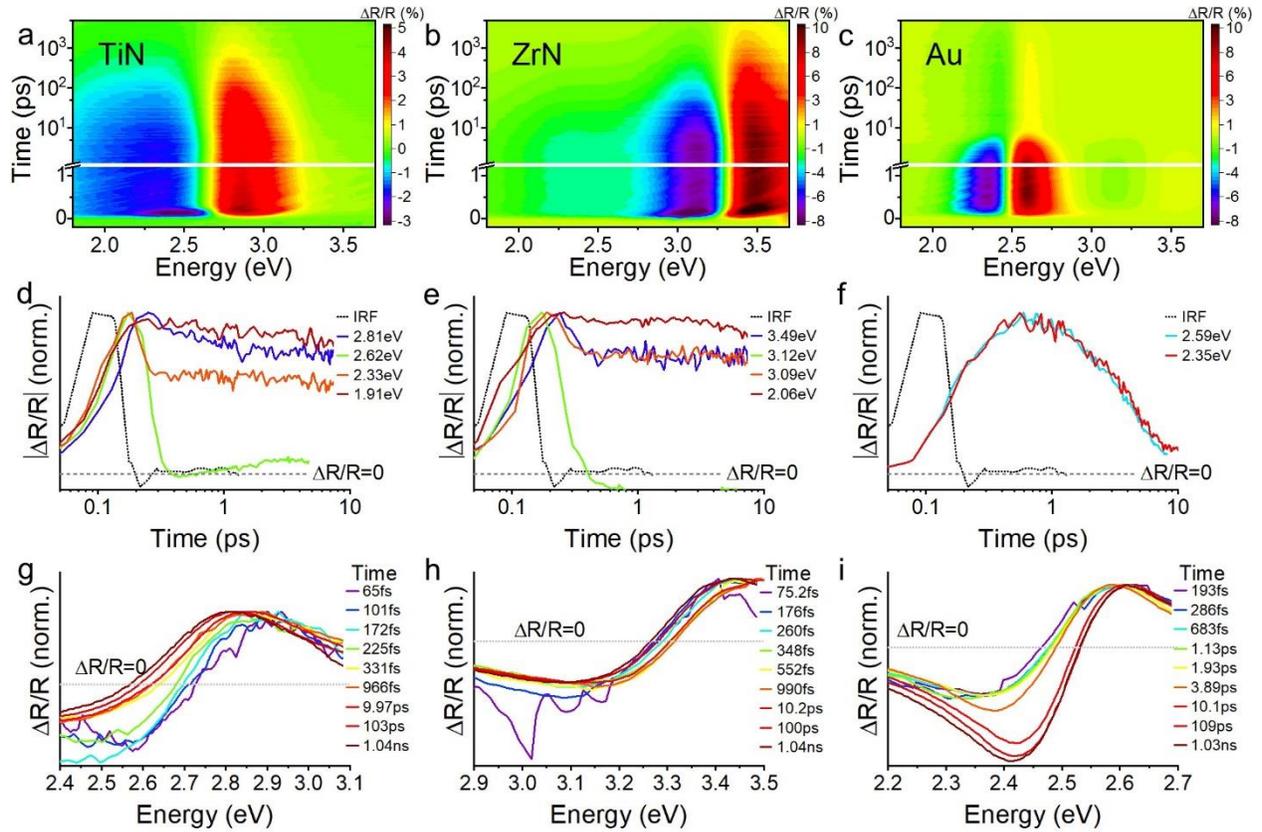

Figure 2. Broadband transient reflection measurements of (a) TiN, (b) ZrN, and (c) Au films under 3.1 eV pump excitation. (d-f) Absolute value of normalized early-time reflectance dynamics for the specified probe energies for the same (d) TiN, (e) ZrN, and (f) Au films. The dashed black line indicates the optical response of glass, estimating the instrument response function (IRF) of the measurement. (g-i) Transient reflection spectra normalized at the peak positive reflectance change for different delay times near the ENZ region for (g) TiN, (h) ZrN, and (i) Au films.

At least three physical processes (in order of typical time scale) are commonly interpreted in the transient response of metallic materials:[69] electron-electron scattering, electron-phonon coupling, and phonon-phonon coupling of the samples to the environment or heat diffusion.[69] Any signature of dephasing of the surface plasmons, often extracted from the homogeneous band-width of nanoparticles,[70,71] is not observed because it occurs in an estimated ~5 fs, much faster than the 35 fs laser pulse.[32,72] Electron-electron scattering is manifest in the rise of signals in Figures 2d, 2e, and 2f. TiN and ZrN show rise times (from 10 % to 90 % of the peak) of ~115 fs and ~105 fs, respectively, whereas Au shows a much longer (~925 fs) rise, consistent with earlier reports.[49,73–75] (Table S1 records values at each displayed energy.) The faster rise times of ΔR/R in TiN and ZrN reflect more rapid electron equilibration in the refractory nitrides compared to Au.

Following electron-electron scattering, the electron plasma cools through electron-phonon coupling, which is observed temporally in Figures 2d-2f and spectrally in Figures 2g-2i. In Au, electron-phonon scattering occurs on a time scale of a ~1 picosecond, as in Figure 2f.[49,51,69,76] The nature of electron-phonon coupling in TiN is debated:[39–42] previous studies probing the metallic spectral window (< 2 eV) showed no sub-picosecond response, which led to contradictory interpretations. One explanation was that electron-phonon coupling in TiN is exceptionally weak,[39] and that there is no sub-picosecond feature because the slow observed dynamics reflect slow cooling of hot carriers. Another hypothesis is that electron-phonon coupling is so fast that the lattice heats up considerably in a sub-picosecond time scale,[42,77] similar to that in transition metals like molybdenum, chromium, or ruthenium.[78] The probe wavelength plays a critical role in the observed dynamics in the first picosecond: TiN and ZrN show sub-picosecond decay features near the ENZ but not in the metallic region. (Figure S3 shows the longer response at the same energies.) The rapid dynamics and spectral changes in this work are consistent with the second hypothesis: the sub-picosecond optical response observed in the ENZ window indicates that TiN and ZrN have strong electron-phonon coupling.

The spectral behavior of TiN, ZrN, and Au are shown in Figures 2g-2i. TiN shows a sub-picosecond redshift whereas ZrN displays a very small sub-picosecond blueshift; both nitrides redshift at longer delay times. The observed transient reflectivity signals may result from several competing phenomena including changes in carrier effective mass,[79,80] population and oscillator strength of allowed interband transitions,[49,61,81] and (at longer delay times) lattice heating and expansion. Focusing on the ENZ window where ΔR/R signals are strongest in both static the optical response is dominated by interband transitions and the transient spectral response is dictated by the density of states at the Fermi level.[52–54,56,62,81,82] Similar to elevated equilibrium temperatures, heating of the electron plasma redistributes electrons through the density of states, changing the spectral distribution of interband oscillator strength. The sub-picosecond spectral changes of TiN and ZrN result principally from the carrier plasma cooling through coupling to phonons. The opposing red-shift of TiN and blue-shift of ZrN are most likely due to distinct band structure and oscillator strength of specific transitions close to the Fermi level, which are not experimentally known,[83] with a potential secondary contribution from residual non-thermal energy distribution of electrons which occurs in the first 100 fs following excitation.[84]

Finally, at longer delay times, the ΔR/R signal for all samples decreases due to cooling of the heated sample back to room temperature through diffusion (thermal conductivity) or radiation of heat. The red-shifting spectral behavior at longer delay times is consistent with lattice cooling, based upon previously presented static changes. Lattice cooling time imposes critical limitations on the use of TiN and ZrN in optical switching.[85] Earlier measurements of TiN found biexponential decays of transient reflectance at longer delay times, a finding repeated here for both TiN and ZrN, with characteristic time scales of tens and hundreds of picoseconds.[39–42] The ΔR/R spectrum of TiN and ZrN subsequent to electron-phonon coupling is a larger fraction of the peak optical response when compared with Au, which is consistent with strong electron-phonon coupling that heats the metal nitride lattice. The persistent, substantial elevation of the lattice and electrons above ambient temperature suggests that these metallic nitrides may be useful for applications in which local heating, such as photothermal therapies or even photocatalysis, are advantageous.

4. Fluence- and Pump Wavelength-Dependent Response

The transient response of TiN and ZrN films is weakly sensitive to pump fluence except for early-time dynamics, which provide additional support for the hypothesis that electron-phonon coupling in transition metal nitrides is rapid. Normalized dynamics at representative energies are shown in Figure 3, with energies chosen to represent close to the maximal positive and negative changes of reflectivity near the ENZ and the metallic region at lower energy. At the representative probe energies, the dynamics of both TiN and ZrN in Figures 3a and 3b show very similar responses at different fluences. This is in contrast with Au, which both in Figure 3c and in many literature reports, shows elongation of the initial decay feature attributed to slowing of electron-phonon coupling.[51,69,86–88] The kinetic data in Figures 3a-3c was fit by biexponential (Au, lowest probe energy TiN and ZrN) or triexponential fits (closer to the ENZ for TiN and ZrN), with fits shown in Figure S4 and fit data presented in Figures 3d-3f. For Au, the initial decay at energies near the intraband absorptions, due to electron-phonon coupling, elongates from 1.4 ps at the lowest fluence to 6.0 ps at the highest fluence. The sub-picosecond decay of TiN and ZrN has an exponential decay time close to or less than 100 fs in most cases. (See also Figure S5.)

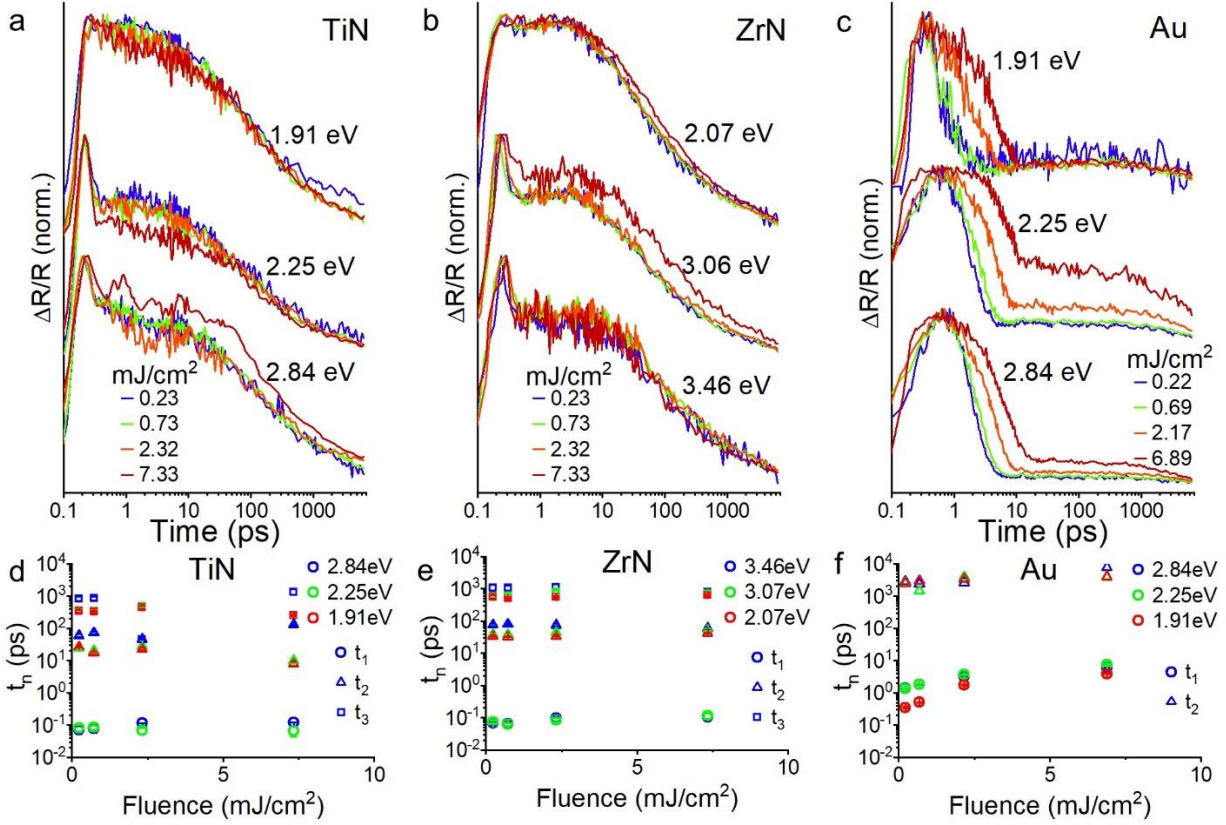

Figure 3. Normalized, fluence-dependent transient reflectivity with 3.1 eV pump photon energy for (a) TiN on MgO, (b) ZrN on MgO, and (c) Au on silicon. (f) Triexponential fit results for the data collected on TiN as a function of fluence. (e) Triexponential fit results for data collected on ZrN as a function of fluence. (f) Biexponential fits results for the data collected on Au as a function of fluence at the specified wavelengths.

The electron-phonon coupling coefficient may be estimated from two-temperature models. Although more complex models invoking hot electron transport[39] or lattice relaxation[42] have been used, the two-temperature model is a reasonable approximation for the data at early delay times for probe energies near the interband transitions. In a two-temperature model, the electrons (at $T_e$) and lattice ($T_l$) reach equilibrium according to the electron-phonon coupling coefficient ($G$) and the respective specific heat capacities of electrons ($C_e$) and the lattice ($C_L$):

$$C_e \frac{\partial T_e}{\partial t} = -G(T_e - T_L) \qquad (2)$$

$$C_L \frac{\partial T_L}{\partial t} = G(T_e - T_L) + G_L(T_L - T_0) \qquad (3)$$

The $G_L$ term reflects phonon-phonon cooling of the lattice. The electron-phonon coupling time, $t_{ep}$, may be approximated as

$$t_{ep}^{-1} \approx G\left(\frac{1}{C_e} + \frac{1}{C_L}\right) \quad (4)$$

which is dominated by the electron heat capacity term. Using the Debye approximation that $C_e = \gamma T_e$ and experimentally measured values of electronic contribution term $\gamma$,[89,90] electron heat capacities of TiN (8.5×10⁴ J·m⁻³·K⁻¹ at 295 K) and ZrN (7.2×10⁴ J·m⁻³·K⁻¹ at 295 K) can be used to estimate a reasonable range of $G$ values. The "zero fluence" value of $t_{ep}$ is estimated by a linear extrapolation of the fluence-dependent fits, shown in Figure S5. Estimates of $G$ are (0.8-1.0)×10¹⁸ W·K⁻¹·m⁻³ for TiN and (1.0-1.02)×10¹⁸ W·K⁻¹·m⁻³ for ZrN, which are close to previous estimates for TiN.[42] These $G$ values are also 25-100 times larger than Au, which is reported as (1.3-4.0)×10¹⁶ W·K⁻¹·m⁻³.[91–96] Thus, in terms of their dynamic response, TiN and ZrN are much more like platinum, which shows a sub-picosecond response in pump-probe measurements, than Au.[78]

Transient reflectivity dynamics at longer times for the TiN and ZrN films are also insensitive to the pump fluence. The decay at longer delay times (> 2 ps) consists of two exponential components, with time constants of the faster component of ~20 ps for TiN and ~35 ps for ZrN followed by a slower time constant of ~350 ps and ~550 ps, respectively. The ΔR/R signal decays over these time scales due to diffusion or radiation of heat from the measured sample spot, including through available interfaces. The substrate's influence on cooling was examined by studying similarly thick TiN and ZrN on silicon. Although electronically different to the TiN and ZrN grown on MgO substrates (ellipsometry in Figure S6), the lattice cooling of the samples shown in Figure S7 is nearly the same. Because samples with different interfaces show similar cooling and, due to acoustic mismatch, heat loss to air is unlikely, the slower biexponential decays of TiN and ZrN are attributed to thermal diffusion within the metal nitride.

ΔR/R increases at higher fluence for both TiN and ZrN, shown in Figure 4a and 4b, respectively. (See also Figure S8.) The increase in ΔR/R with fluence is approximately linear, with a peak response of 15 % for TiN and 52 % for ZrN with 3.10 eV excitation at a fluence of 7.33 mJ·cm$^{-2}$. More critically, the spectral responses at early time provide additional evidence in support of sub-picosecond electron-phonon coupling. As shown in Figures 4c and 4d, the band-width over which sub-picosecond dynamics occurs grows with higher fluence. For example, a higher fluence must be used to observe a measurable sub-picosecond component in the transient reflection dynamics of TiN at 2.1 eV or ZrN at 2.5 eV. This phenomenon arises because as electrons are driven to higher effective temperatures by increasing pump fluence, the electron energy distributions smears over more states, which broadens photoinduced interband transitions.[81] Measurements at low energy (< 2 eV) can only reveal the sub-picosecond dynamics indicative of strong electron-phonon coupling at high fluence.

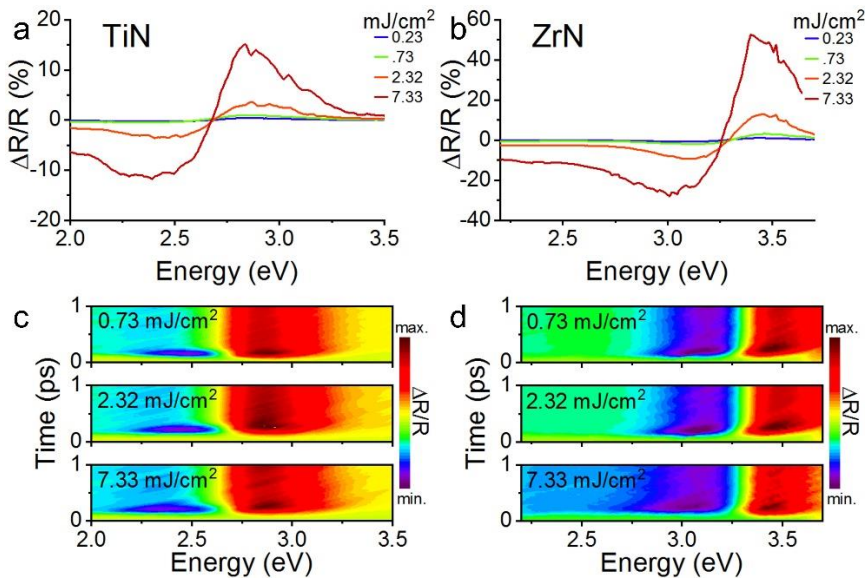

Figure 4. Fluence-dependent spectra of (a) TiN and (b) ZrN films at 200 fs pump-probe delay for the specified fluences. (c, d) Two-dimensional plots of the first picosecond delay shown for (c) TiN and (d) ZrN films at selected fluences. The color scales of the data are normalized to the minimum and maximum within the map window.

The values of ΔR/R at longer delays (> 2 ps) permit estimates of lattice heating based upon static thermoreflectance coefficients, although agreement between static and transient ΔR/R is imperfect (Figure S9). In addition to the fast dynamics at early time, the other important consequence of strong electron-phonon coupling in TiN and ZrN is substantial heating of the lattice compared to Au. For example, Figure S10 shows the lattice heating of ZrN to 405 K, TiN to 325 K, and Au to 315 K with a fluence close to 2.3 mJ·cm$^{-2}$ (3.10 eV photons). The greater lattice heating in TiN and ZrN emphasizes the potential of these materials for photothermal applications and this result is consistent with recent findings of photothermal transduction efficiency of ZrN > TiN > Au.[24]

The pump photon energy was also varied to study the ultrafast response. Unlike in the case of Au,[97] the dynamics of TiN and ZrN samples displayed little sensitivity to pump photon energy, as shown in Supporting Information Figure S11. However, the absolute amplitude of reflectivity changes is sensitive to pump photon energy. As shown in Figure 5a, the maximal change in reflectivity for both TiN and ZrN increases with the pump photon energy, apart from a dip from trend in TiN at 3.1 eV, which is a region of increasing reflectivity and minimized losses (Figures 1a and 1b). The increase in differential reflectivity at higher pump photon energies is a consequence of the larger thermal dissipation (loss) and lower reflectivity due to interband absorption. Lattice heating under these excitation conditions is estimated in Figure 5b, conveying the larger heating achieved with blue photoexcitation: both TiN and ZrN reach lattice temperatures above 400 K for 3.89 eV photoexcitation at a fluence of 2.3 mJ·cm$^{-2}$. The energy-dependent response has clear implications in photothermal applications, where TiN and ZrN will likely show a strong spectral sensitivity in heating. Excitation on resonance with interband transitions drives large increases in heating and those at low energies requires greater input fluence to drive similar lattice heating.

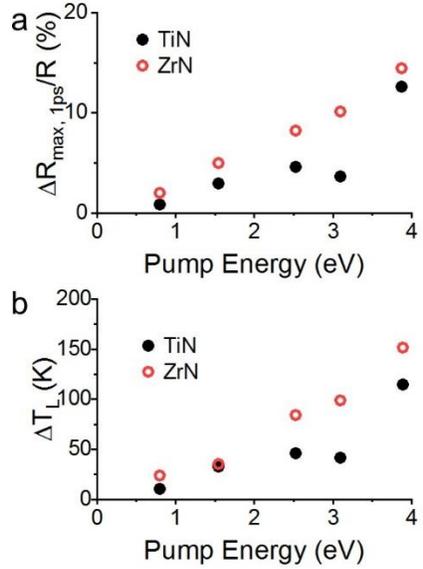

Figure 5. (a) Maxima of transient reflectivity spectra versus pump photon energy. (b) Estimated temperature rise of the samples under the given excitation conditions. Except for 0.80 eV (12 mJ/cm$^2$), the pump fluence was kept at 2.3-2.4 mJ/cm$^2$.

6. Conclusion

This work reports broadband ultrafast reflectance measurements of metallic TiN and ZrN thin films spanning the ultraviolet and visible wavelength regions and a range of pump fluences and pump photon energies. The measurements reveal a sub-picosecond response of the materials in the epsilon-near-zero window attributable to a short-lived electron energy redistribution which changes the spectral character of interband transitions. These measurements resolve previous ambiguity regarding whether electron-phonon coupling in these materials is strong or weak: Both materials are found to exhibit strong electron-phonon coupling, 25-100 times greater than Au. Strong electron-phonon coupling results in cooling of electrons on time-scales less than 100 fs and greater heating of the TiN and ZrN lattice than observed for Au under comparable conditions. The larger thermal response of the TiN and ZrN lattice under photoexcitation supports previous findings of the promise the use of these materials in photothermal applications.

7. Experimental Section

*Fabrication*: The titanium nitride films were deposited on magnesium oxide substrates using DC magnetron sputtering at 800 °C. A 99.995% pure titanium target of 2 in. diameter was used. The DC power was set at 200 W. To maintain a high purity of the grown films, the chamber was pumped down to $3\times10^{-8}$ Torr prior to deposition and backfilled to 5 mTorr during the sputtering process with argon. The throw length was kept at 20 cm, ensuring a uniform thickness of the grown TiN layer throughout the 1.5 cm by 1.5 cm sapphire substrate. After heating, the pressure increased to $1.2\times10^{-7}$ Torr. An argon–nitrogen mixture at a rate of 4 sccm/6 sccm was flowed into the chamber. The deposition rate was 2.2 Å min$^{-1}$. For TiN on silicon, the conditions were identical, except that an argon: nitrogen gas ratio of 1sccm/18sccm was used. The growth rate was around 1.5 nm min$^{-1}$. On MgO, the TiN films showed epitaxial growth. On silicon, the TiN films grown were polycrystalline in nature. For ZrN growth, the substrate was heated to a temperature of 850 °C. A 99.95 % pure zirconium target was used, and the argon-nitrogen mixture was flown at a rate of 1sccm/18sccm. The remaining growth parameters were identical to that of titanium nitride. The deposition rate was 3.0 nm min$^{-1}$ for both MgO and silicon substrates. The ZrN on both MgO and silicon are polycrystalline in nature.

*Static Optical Measurements:* Normal incidence reflectivity was taken using a Filmetrics F40 optical profilometer with a standard calibration wafer to ensure accurate absolute reflectivity. Temperature-dependent measurements of reflectivity, from which were calculated differential reflectance spectra, were collected with a halogen lamp and an OceanOptics USB2000 spectrometer. A silver mirror was used as a background with reflectivity corrected for the wavelength-dependent reflectivity of silver. Samples were mounted in a Janis VPF 800 cryostat with calcium fluoride windows and data was collected at a 60° total angle to match the angle of transient measurements. UV-NIR ellipsometry was performed from 300 nm to 2300 nm using a J. A. Woollam V-Vase UV-vis-NIR spectroscopic ellipsometer, at incident angles of 50° and 70°.

The data was fit using a Drude-Lorentz model, comprising one Drude term and two Lorentz terms, an approach which is common for these materials.[55] The Drude term models the free carrier plasma and the Lorentz terms model interband transitions. Fit data are shown in Figure S12 with strong agreement to raw spectroscopic ellipsometry over the energy range of interest. The relative permittivity can be represented as a complex number, given by the following equation,

$$\varepsilon = \varepsilon_1 + i\varepsilon_2 = \varepsilon_{1\infty} + \sum_k \frac{B_k}{E_k^2 - (h\nu)^2 - iC_k h\nu} \qquad (5)$$

where $\varepsilon_1$ and $\varepsilon_2$ are the real and imaginary parts of the dielectric constant. $B_k$ is the amplitude, $C_k$ is the broadening and $E_k$ is the center energy for the $k$-th oscillator; $h\nu$ is the incident photon energy in eV, and $\varepsilon_{1\infty}$ is an additional offset term defined in the model. $B_0$ and $C_0$ are terms corresponding to the Drude term, with $E_0 = 0$.

*Time-Resolved Optical Measurements:* Ultrafast transient reflectivity was collected using the split output of an 800 nm 35 fs Ti: sapphire. A chopped pump excitation employed either the fundamental (800 nm) or second harmonic (400 nm) of the Ti: sapphire laser, or used an optical parametric amplifier to generate other pump beams energies. A delayed white light supercontinuum probe was generated by focusing the 800 nm laser into a calcium fluoride plate. The pump beam was oriented with an angle of incidence close to 20°; the probe beam was oriented with an angle of incidence of ~30° (60° total angle).

Acknowledgements

This work was performed, in part, at the Center for Nanoscale Materials, a U.S. Department of Energy Office of Science User Facility, and supported by the U.S. Department of Energy, Office of Science, under Contract No. DE-AC02-06CH11357. This work as also supported, in part, by the Air Force Office of Scientific Research through awards numbered FA9550-17-1-0243, FA9550-18-1-0002, and FA9550-19-S-0003.

(94) Brorson, S. D.; Kazeroonian, A.; Moodera, J. S.; Face, D. W.; Cheng, T. K.; Ippen, E. P.; Dresselhaus, M. S.; Dresselhaus, G. Femtosecond Room-Temperature Measurement of the Electron-Phonon Coupling Constant in Metallic Superconductors. *Phys. Rev. Lett.* **1990**, *64* (18), 2172–2175. https://doi.org/10.1103/PhysRevLett.64.2172.

(95) Elsayed-Ali, H. E.; Juhasz, T.; Smith, G. O.; Bron, W. E. Femtosecond Thermoreflectivity and Thermotransmissivity of Polycrystalline and Single-Crystalline Gold Films. *Phys. Rev. B* **1991**, *43* (5), 4488–4491. https://doi.org/10.1103/PhysRevB.43.4488.

(96) Choi, G.-M.; Wilson, R. B.; Cahill, D. G. Indirect Heating of Pt by Short-Pulse Laser Irradiation of Au in a Nanoscale Pt/Au Bilayer. *Phys. Rev. B* **2014**, *89* (6), 064307. https://doi.org/10.1103/PhysRevB.89.064307.

(97) Minutella, E.; Schulz, F.; Lange, H. Excitation-Dependence of Plasmon-Induced Hot Electrons in Gold Nanoparticles. *J. Phys. Chem. Lett.* **2017**, *8* (19), 4925–4929. https://doi.org/10.1021/acs.jpclett.7b02043.

Supplementary Information

# Broadband Ultrafast Dynamics of Refractory Metals: TiN and ZrN

*Benjamin T. Diroll,[1]\* Soham Saha,[2] Vladimir M. Shalaev,[2] Alexandra Boltasseva,[2] and Richard D. Schaller[1,3]*

[1]Center for Nanoscale Materials, Argonne National Laboratory, Lemont, Illinois 60439

[2]School of Electrical and Computer Engineering, Purdue University, West Lafayette, Indiana 47907

[3]Department of Chemistry, Northwestern University, 2145 Sheridan Road, Evanston, Illinois 60208

[*bdiroll@anl.gov](*bdiroll@anl.gov)

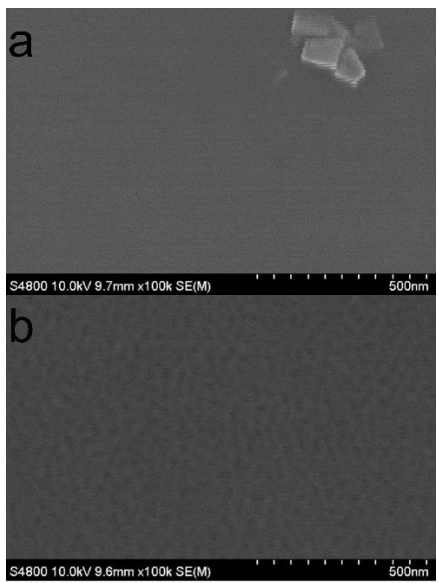

Figure S1. Scanning electron micrograph images of (a) TiN and (b) ZrN films on MgO. In (a), the large particles on the surface are not representative of the average surface, but they are included to convey focus and the mostly featureless film morphology of TiN (a) at this magnification, compared to ZrN (b).

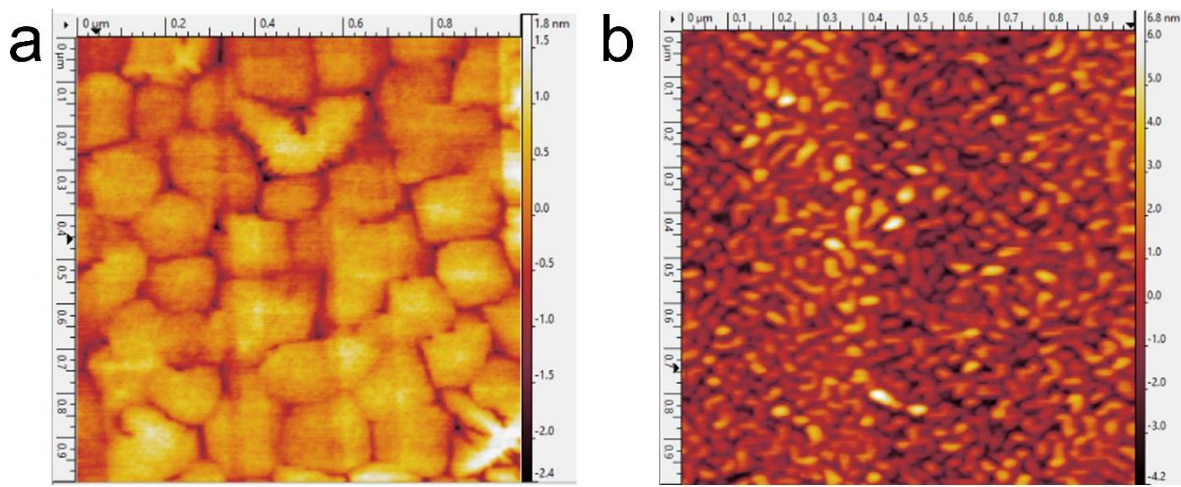

Figure S2. Atomic force microscopy contour plots of (a) TiN and (b) ZrN films on MgO.

Table S1. Approximate Rise Times of Samples at Specified Probe Energies using 3.1 eV Pump Excitation.

| Sample | Probe Energy (eV) | Rise Time (fs, 10% to 90%) |
| --- | --- | --- |
| TiN | 2.81 | 129 |
|  | 2.62 | 100 |
|     | 2.33 | 100 |
|     | 1.91 | 125 |
| ZrN | 3.49 | 137 |
|     | 3.12 | 79  |
|     | 3.09 | 90  |
|     | 2.06 | 115 |
| Au  | 2.59 | 951 |
|     | 2.35 | 903 |

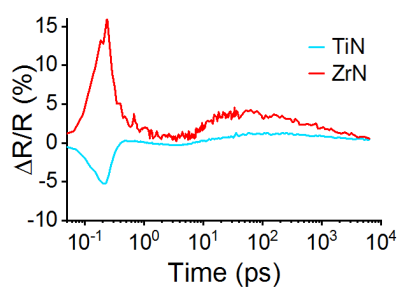

Figure S3. Kinetics of TiN and ZrN near the ENZ point for 3.10 eV pump photon energy with a fluence of 7.33 mJ·cm$^{-2}$.

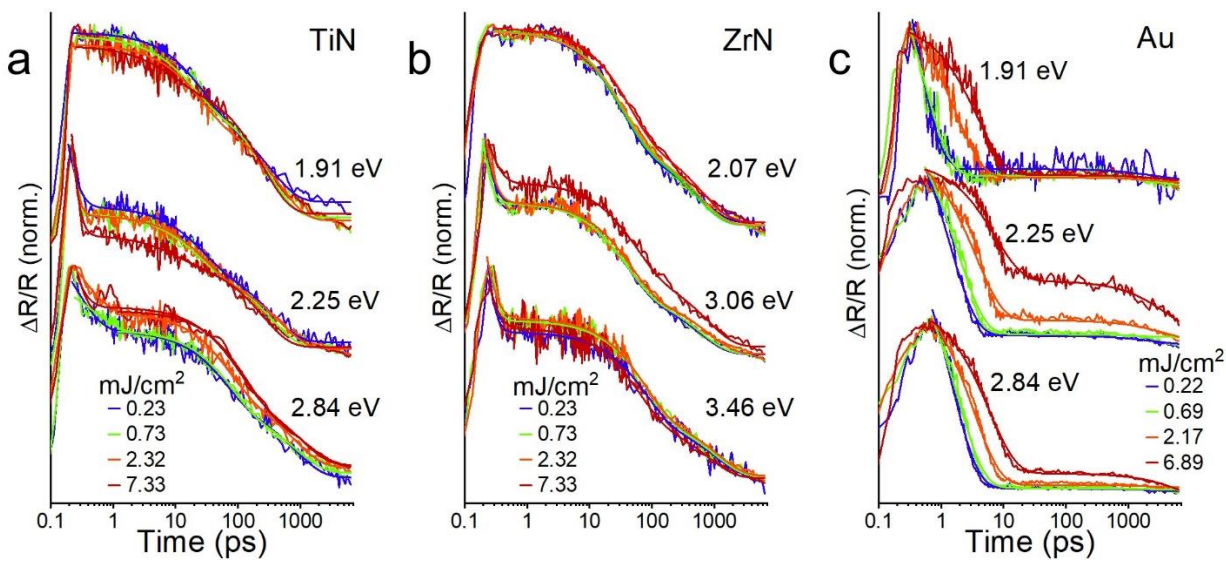

Figure S4. Fluence-dependent dynamics for 3.1 eV pump photon energy excitation of (a) TiN and (b) ZrN films on silicon and (c) Au film on silicon. Data are shown for selected probe energies shown on right. Solid fit lines are overlaid on the data.

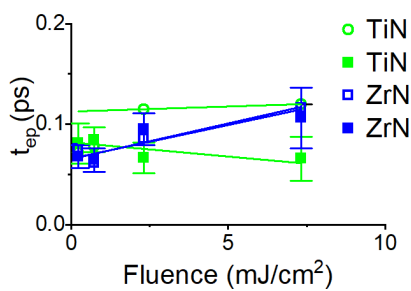

Figure S5. Linear fits to electron-phonon coupling time constant ($t_{ep}$, symbols) with variable fluences. Open symbols represent the higher probe and closed symbols the lower probe energy. Data from Figures 3 and S4. The y-intercept is taken as the fundamental electron-phonon coupling time constant of the material.

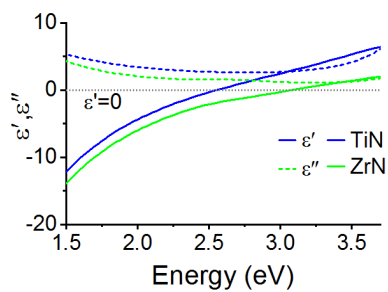

Figure S6. Ellipsometry results of TiN and ZrN films grown on silicon.

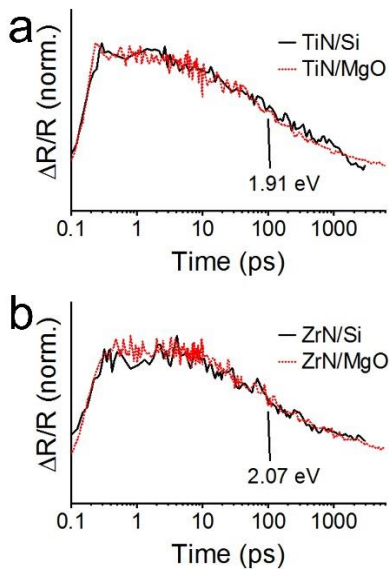

Figure S7. Normalized transient reflectivity under 3.89 eV pump photon energy excitation of (a) films of TiN on silicon and MgO and (b) films of ZrN on silicon and MgO. The probe

wavelength for the measurement is shown in the plot; it was chosen for the reflective region, away from interband absorptions, which are somewhat different in the two samples.

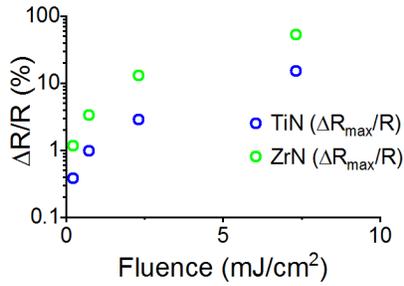

Figure S8. Peak differential reflectivity of TiN and ZrN at 200 fs delay following 3.10 eV pump photon energy excitation at the given fluences.

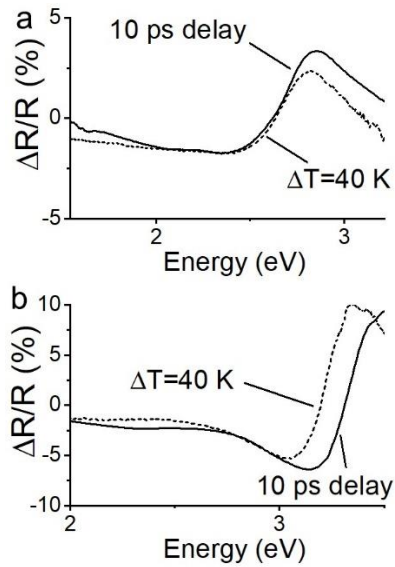

Figure S9. (a) Raw differential reflectivity data of TiN film on MgO from static thermoreflectance measurements (dashed) with a temperature difference of 40 K (starting from 295 K) and the transient reflectivity spectrum at 10 ps delay for the same film with 3.1 eV photon excitation. (b) analogous data for ZrN film on MgO.

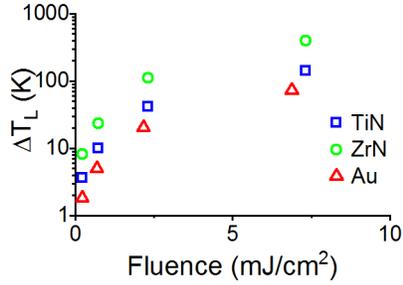

Figure S10. Estimated lattice temperature rise after 2 ps (for TiN and ZrN) and 20 ps (for Au) plotted against fluence for 3.10 eV pump photon energy. Estimates of the lattice temperature rise for fluences for 2.3 mJ·cm$^2$ and below are interpolations based upon the linear regime of static temperature-dependent reflectance. The highest fluence may fall outside of the linear response regime and therefore most likely an overestimate of the temperature rise. Consequently, values at 2.3 mJ·cm$^2$ are the highest temperature changes reported in the text.

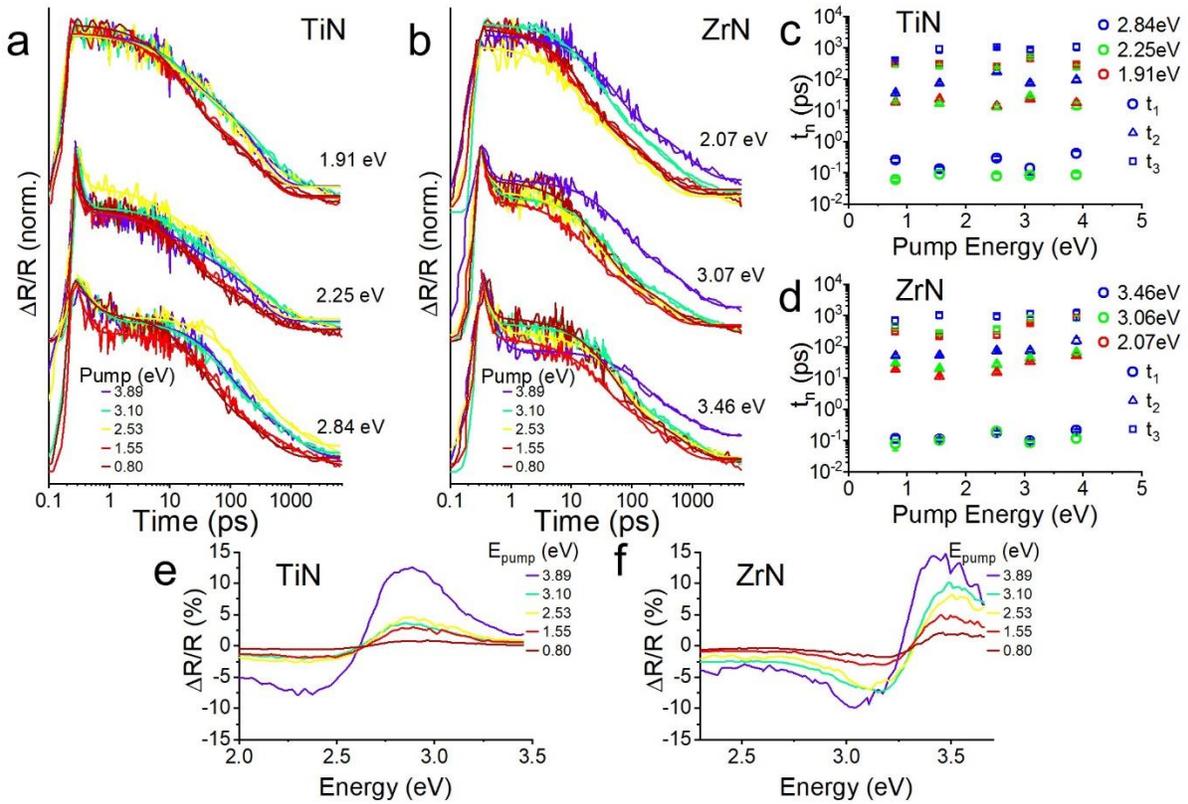

Figure S11. Dynamics of (a) TiN and (b) ZrN films on MgO for specified pump photon energies, measured at probe wavelength specified at right on plots. Solid fit lines are overlaid on data. Fit results for (c) TiN and (d) ZrN are shown in the scatterplots. Spectra of (e) TiN and (f) ZrN at 1 picosecond delay for specified pump photon energies. Except for 0.80 eV (12 mJ/cm$^2$), the pump fluence was kept at 2.3-2.4 mJ/cm$^2$.

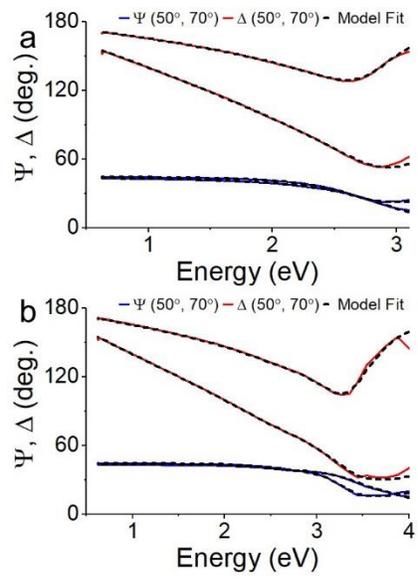

Figure 12. Raw spectroscopic ellipsometry results (Ψ and Δ) shown in solid lines of blue and red, respectively, with a dashed line showing the model fit of the data for (a) TiN on MgO and (b) ZrN on MgO.